\documentclass[3p,times]{elsarticle}

\usepackage{ecrc}
\usepackage{color}

\volume{00}

\firstpage{1}

\journalname{Icarus}

\runauth{}

\jid{icarus}

\jnltitlelogo{ICARUS}

\CopyrightLine{2011}{Published by Elsevier Ltd.}

\usepackage{amssymb}
\usepackage[figuresright]{rotating}

\begin{document}

\begin{frontmatter}



\dochead{}
%% e.g. \dochead{17th International Conference on Dynamical Processes in Excited States of Solids}

\title{Thermal Infrared and Optical Photometry of
  Asteroidal Comet C/2002~CE$_{10}$}


\author{Tomohiko Sekiguchi{\rm{$^{a}$}}}
\author{Seidai Miyasaka{\rm{$^{b}$}}}
\author{Budi Dermawan{\rm{$^{c}$}}}
\author{Thomas Mueller{\rm{$^{d}$}}}
\author{Naruhisa Takato{\rm{$^{e f}$}}}
\author{Junichi Watanabe{\rm{$^{e}$}}}
\author{Hermann Boehnhardt{\rm{$^{g}$}}}

\address{{\rm $^{a}$} %Department of Teacher Training, 
Hokkaido University of Education, 9 Hokumon, Asahikawa, Japan}
\address{{\rm $^{b}$} Tokyo Metropolitan Government, 2-8-1, Nishishinjuku, Shinjuku, Tokyo, Japan}
\address{{\rm $^{c}$} Department of Astronomy, Bandung Institute of Technology, Bandung 40132, Indonesia}
\address{{\rm $^{d}$} Max-Planck-Institute for Extraterrestrial Physics, Giessenbachstrasse, 85748 Garching, Germany}
\address{{\rm $^{e}$} National Astronomical Observatory of Japan, 2-21-1  Osawa, Mitaka, Tokyo, Japan}
\address{{\rm $^{f}$} Subaru Telescope, 650 North A'ohoku Pl., Hilo, HI, USA}
\address{{\rm $^{g}$} Max-Planck-Institute for Solar System Research, Justus-von-Liebig-Weg 3, 37077 G\"ottingen, Germany}

\begin{abstract}
C/2002~CE$_{10}$ is an object in a retrograde elliptical orbit with Tisserand parameter  $-0.853$
indicating a likely origin in the Oort Cloud. It appears to be a rather inactive comet since no coma 
and only a very weak tail was detected during the past perihelion passage.
We present multi-color optical photometry,
lightcurve and thermal mid-IR observations of the asteroidal comet.
With the photometric analysis in $BVRI$, 
the surface color is found to be redder than asteroids,
corresponding to cometary nuclei and TNOs/Centaurs. 
The time-resolved differential photometry supports
a rotation period of 8.19$\pm$0.05~h.
The effective diameter and the geometric albedo are 
17.9$\pm$0.9~km and 0.03$\pm$0.01,
%$\sim$18~km and 0.03, 
respectively,
indicating a very dark reflectance of the surface.
The dark and redder surface color of C/2002~CE$_{10}$ may be attribute 
to devolatilized material by surface aging suffered from the irradiation 
by cosmic rays or from impact by dust particles in the Oort Cloud.
Alternatively, C/2002~CE$_{10}$ was formed of very dark refractory material
originally 
like a rocky planetesimal.
In both cases, this object lacks ices (on the surface at least).
The dynamical and known physical characteristics of C/2002~CE$_{10}$ are best compatible
with those of the Damocloids population in the Solar System, that appear to be exhaust
cometary nucleus in Halley-type orbits.
The study of physical properties of rocky Oort cloud objects 
may give us a key for the formation of the Oort cloud and the solar system.
\end{abstract}

\begin{keyword}


Asteroids \sep
Asteroids surface \sep
Comets \sep
Comets nucleus \sep
Infrared observations \sep
Photometry \sep 
\end{keyword}

\end{frontmatter}

% \linenumbers

\section{Introduction}
\label{intro}

On February 2, 2002, the LINEAR project of the MIT Lincoln Laboratory in Socorro, USA, discovered 
an object approaching the Sun at about Jupiter's distance. Although of asteroidal appearance without 
sign of activity, it was classified as cometary object, C/2002~CE$_{10}$, based upon its retrograde 
comet-like orbit (see Table.~\ref{orbit}). 
Deep imaging of the object, obtained with the Subaru telescope
around the period of closest approach to Earth, 
revealed a short faint tail of C/2002~CE$_{10}$ (Takato et al. 2003). 
The faint tail may either be caused by very weak or by temporal cometary activity
(sublimation of gas and release of embedded dust), or it may be due to a recent impact event of low, but non-zero 
occurrence probability.
Since cometary coma activity in C/2002~CE$_{10}$ has not been reported so far despite deep imaging attempts 
using the Subaru telescope's Prime Focus Camera: Suprime-Cam,
%\textcolor{blue}{(Takato in preparation)} %(see Fig.~\ref{PSF})
and despite the object passed through perihelion well within the water sublimation limit, 
C/2002~CE$_{10}$ may represent a transitional object between the population classes of comets and asteroids.
This paper primarily presents an analysis of the nucleus properties of this object
and the properties of dust tail with the Finson-Probstein
analysis and no coma activity will be analyzed in a future paper.

%__________________________________________________ One column table
\begin{table}
\begin{center}
 \caption{\label{orbit} Orbital information of C/2002~CE$_{10}$ (orbital elements taken from M.P.E.C. 2003-R41}
\begin{tabular}{l|l}
\hline
\hline
$a$: semimajor axis  & 9.815 (au)\\
$q$: perihelion distance & 2.047 (au)\\
$Q$: aphelion distance & 17.585 (au)\\
$e$: eccentricity 	& 0.7915 \\
$i$: inclination 	& 145.46 ($^{\circ}$) \\ 
$\omega$: argument of perihelion & 126.19 ($^{\circ}$) \\
$\Omega$: longitude of the ascending node  & 147.44 ($^{\circ}$) \\
$M$: mean anomaly & 0.0609 ($^{\circ}$) \\
$n$: mean motion & 0.0320 ($^{\circ}$/d) \\
%$\bf HBO note:$           & enter full set of orbital elements plus Tisserand parameter in this table\\
\hline
Perihelion Passage & 2003, June 22.10 TT \\
Earth Approach & 2003, Sept 04.90 TT ({$\mit  \Delta$} = 1.231 au) \\
$P$: orbital period 	& 30.75 (years)\\
\hline
$T_{\mathrm J}$: Jupiter Tisserand parameter & -0.853\\
\hline
\end{tabular}
\end{center}
\end{table}
%__________________________________________________________________

%\begin{figure}
%   \centering
%\includegraphics[width=9cm,clip]{PSFcomparison.eps}
%\caption{Comparison between the radial profile of C/2002~CE$_{10}$ (solid line) 
%and the point-spread-function (PSF) of a field star (broken line) in the comet exposures. 
%Comet and PSF star profiles were extracted from deep imaging exposures of C/2002~CE$_{10}$ 
%in $R$-band, performed with the Subaru telescope in 2002, 16 and 17 August
%as well as on 2002, 22 and 23 September, using Suprime-Cam (Takato et al. 2003).
%The radial profile of C/2002~CE$_{10}$ matches well that of the PSF star.
%Note: A very faint straight tail of about 21" extension is noticeable in the {bf HBO: you may want to add the date of the observations} August 2002 
%at position angle (PA) 212~deg.} 
%\label{PSF}
%\end{figure}

%%
%%
%%
%%
%%

The criteria for classification a minor body as asteroid
or as comet are 
appearance (coma and tail versus point-like) and 
orbit parameters, namely the Tisserand parameter; $T_{\mathrm J}$.
%A quantitative definition of Halley-type objects coming from
%the Oort Cloud is made using the Tisserand invariant $T$  (e.g. Carusi \& Valsecchi, 1987). 
The parameter $T_{\mathrm J}$ characterizing the dynamical link of minor
bodies to the gravitational disturbance by planet Jupiter (Carusi et al. 1995) is used 
to differentiate the Halley-type comets 
($T_{\mathrm J}<2$) from the Jupiter-family comets
($3>T_{\mathrm J}>2$) and 
objects with $T_{\mathrm J}>3$ are generally considered
to be dynamically asteroidal.
%$T_{\mathrm J}$ is given by
It is given by
$
T_{\mathrm J} = {{a_{\mathrm J}} \over {a}} + 
2  \left\{{ {{a_{\mathrm J}} \over {a}} (1 - e^2) } \right\}^{1 \over 2} 
\cos (i),$ 
where
${a_{\mathrm J}}$ and $ {a}$ are the semi-major axis of Jupiter and the object (asteroid, comet or others)
respectively, $i$ and $e$ are the object's inclination and eccentricity,
respectively. 
%\textcolor{blue}{
%$T_{\mathrm J}$ is 
%employed to examine whether minor bodies in the solar system including asteroids originate
%from the Oort Cloud or from the Kuiper Belt (e.g. Weissman et al, 1989, Weissman et al. 2002).
%}
Recently, this classification approach is challenged
by the discovery of objects of cometary appearance in asteroid-like orbits, the so called ''Main Belt Comets'' (MBCs), 
and of objects of point-like appearance in cometary orbits. 
Jewitt (2005) assigned Halley-type orbit asteroids and inactive comets into a new group, the ``Damocloids'', 
named after asteroid (5335) Damocles. 
%The MBCs may even represent a third reservoir of
%cometary bodies in the Solar System aside the Kuiper Belt and the Oort Cloud.
Meech et al. (2016) reported that
Oort cloud comet C/2014~S3 (PANSTARRS) shows a very weak level of cometary activity and 
S-type asteroid spectra with a silicate absorption feature around 1~$\mu$m wavelength, indicating that 
the comet may be physically similar to an inner main
belt rocky S-type asteroid.

The $T_{\mathrm J}$-value of C/2002~CE$_{10}$ is  $ -0.853$ (Tab. 1),
indicating in object in an Halley-type orbit. 
This result together with no coma appearance (Takato et
al. 2003)
%point-spread-function PSF (Fig. 1), 
suggests that
C/2002 CE$_{10}$ might be either an extinct comet or a quasi-inert object that has been eject from the 
Oort Cloud, giving us a good opportunity to investigate basic physical characteristics of an Oort Cloud 
object. Furthermore, the results may provide insights in the surface aging of minor bodies and the link 
between asteroids and comets. 
After an outline of the observations and data reduction performed, we present results on physical properties of 
C/2002 CE$_{10}$: rotation period, axis ratio, dimension, albedo and color taxonomy. In the final section of
the paper we discuss, based upon our findings, the relation of C/2002 CE$_{10}$ with other minor body 
populations in the solar system.

\section{Observations and Data Reduction}
\label{observations}
C/2002 CE$_{10}$ was observed in September and October 2003 in the visible and thermal infrared
wavelength ranges.
 
\begin{table*}
\caption{\label{Obsdata} Observing Geometry of C/2002~CE$_{10}$}
\begin{tabular}{lcccccc}
\hline
Date & ${R_{\mathrm h}}^{\mathrm a}$ & ${\mit{\Delta}}^{\mathrm b}$ & Phase Angle  &
Observation Type (\& band) & Telescope & sky condition\\
 (UT) & (au) 	     & (au) 		        & (deg.) &  & &\\
\hline
\hline
%2003-Aug-21 & 2.15	& 1.33	& 20.2$^{\circ}$  & imaging ($R$) & Subaru 8.2~m\\
%22 Aug. 2003 & 2.15	& 1.32	& 19.7$^{\circ}$  & imaging ($R$)& Subaru 8.2~m\\
2003-Sep-06 & 2.20	& 1.23	& 9.8$^{\circ}$  & thermal
IR ($N$) & ESO 3.6~m & photometric\\
2003-Oct-02 & 2.31	& 1.57	& 20.1$^{\circ}$ & color ($BVRI$) \& lightcurve
($I$) & Kiso 1.05~m & photometric\\
2003-Oct-03 & 2.32	& 1.59	& 20.5$^{\circ}$ & lightcurve ($I$)& Kiso 1.05~m & thin clouds\\
2003-Oct-04 & 2.32	& 1.61	& 20.9$^{\circ}$ & lightcurve ($I$)& Kiso 1.05~m & thin clouds\\
2003-Oct-06 & 2.33	& 1.66	& 21.6$^{\circ}$ & lightcurve ($I$)& Kiso 1.05~m & thin clouds\\
2003-Oct-08 & 2.34	& 1.71	& 22.2$^{\circ}$ & lightcurve ($I$)& Kiso 1.05~m & thin clouds\\
\hline
\end{tabular}
\begin{list}{}{}
\item[$^{\mathrm{a}}$] $R_{\mathrm h}$ is the heliocentric distance.
\item[$^{\mathrm{b}}$] ${\mit \Delta}$ is the geocentric distance.
\end{list}
\label{ephemtab}
\end{table*}

\subsection{Optical Observations}

The $BVRI$ observations of C/2002~CE$_{10}$ were carried out on a photometric
night, 2003-Oct-3, using the 1.05~m Schmidt telescope at the Kiso observatory, Japan. 
The CCD camera used has 2048 $\times$ 2048 pixels with pixel size of 24 $\mu$m and 
covers a field of view (fov) of 50$'$ $\times$ 50$'$. It is well-suited for time-resolved 
observations of Solar System objects using an adequate number of reference
stars in the fov for differential photometry of moving targets.
$B$ and $V$ exposures were taken through Johnson-type filters, $R$ and $I$ exposures
through Cron-Cousins-type filters. The multi-color photometric observations were embedded 
in the series of $I$-band exposures for lightcurve sampling
(i.e. --$I$--$I$--$B$--$I$--$V$--$I$--$R$--$I$--$I$-- 
sequence) in order to follow and compensate for brightness variations due to non-spherical 
shape or albedo in combination with the rotation motion of the object.
The photometric parameters of the telescope-instrument combination and of the atmosphere were
determined by measuring various standard star fields at
different airmasses 
%{\bf HBO comment: you may want to give  a list of the at different fields used from the Landolt list}. 
Lightcurve observations were performed between
2003-Oct-2 and 2003-Oct-8, occasionally with thin
clouds. 
  In order to minimize the fluctuation of sky conditions and scattering of lunar light by thin cloud, $I$-band filter was used.
The standard calibration frames 
(CCD bias and flatfield exposures) were also obtained as needed.
Table~\ref{ephemtab} summarizes the observing geometry, exposures types and sky 
conditions for the C/2002~CE$_{10}$ observations.

Differential photometry between C/2002~CE$_{10}$ and comparison stars in the fov is applied.
In order to reduce
the influence of possible variable stars on the photometric results, to gain 
a high signal-to-noise ratio, and to ensure the confidence of the measurements and a good
coverage for the lightcurve analysis, we selected comparison stars according to the 
following criteria:
(1) as many as possible, (2) as bright as possible, (3) with
maximum exposure level below 40,000 ADU to stay well inside the linearity range of the CCD
detector, (4) visible and measurable in the whole set of images of a single night. 
Daily extinction and zero-point parameters were derived from the exposures of the Landolt 
standard star fields and by measuring the comparison stars in the object fov. Magnitudes of 
comparison stars are derived per night series using photometric data of the standard stars 
at the same airmasses.

\subsection{Thermal mid-IR Observations}

Thermal observations of C/2002~CE$_{10}$ were carried out on 2003-Sep-6. 
$N$-band images were taken in service mode with the 3.6~m telescope
and the TIMMI2 instrument at the La Silla site of the European Southern Observatory ESO in Chile. 
TIMMI2, the Thermal Infrared Multi-Mode Instrument~2 (K{\" a}ufl et al., 2003) 
has a 240 $\times$ 320 pixel SiAs detector, and it is operated at
6.5-7.5~K. The image scale used for our observations was $0.''202$ ${\mathrm {pixel}}^{-1}$ 
which offers a field of view of $64''\times 48''$ on the sky. The $N$1-filter of the 
TIMMI2 instrument (with effective central wavelength of 8.6~$\mu$m) was chosen because of
an expected advantageous sensitivity for the observations of C/2002~CE$_{10}$. 
The individual TIMMI2 detector integration time (DIT) was set to 20.8 milliseconds. 
The observations were performed as a series of 4 exposures using 
secondary mirror chopping and telescope nodding as follows: 
On target position 3 DIT read-outs were taken at two chopping positions offset in 
North-South direction by $10''$. This
chopping-integration cycle was repeated 60 times. Thereafter, the telescope was
moved $10''$ in East-West direction and 
80 chopping-integrations were repeated as before.
An exposure series of C/2002~CE$_{10}$ were made with a total integration
time of 2396.16~sec.
During the observations of C/2002~CE$_{10}$ the sky conditions were photometric 
with an average seeing of 1$''$.

The basic reduction for C/2002~CE$_{10}$ data makes use the TIMMI2 reduction pipeline (Relke et al. 2000, Siebenmorgen et al., 2004). 
The pipeline procedure
automatically subtracts the pairs of ``chopped'' images 
and co-adds all the frames of the whole chopping/nodding sequence 
(equivalent to one exposure series). 
Hence, the resulting image shows 2 positive and 2 negative 
sub-images of C/2002~CE$_{10}$.
The two negative ones are multiplied by -1 and all sub-images of C/2002~CE$_{10}$ are
shifted such that the pixel positions of the brightness center in the sub-images overlap.
At the end the shifted subimages are coadded to the result frame of the respective
exposure series.
The TIMMI2 data are flux-calibrated using observations
  of standard star HD156277 which
were obtained during the same night applying the same filter setup and observing mode
as for the observations of C/2002~CE$_{10}$. An airmass
correction factor of 7~\%
is applied to
compensate for the different airmasses of the standard star (airmass = 1.29) and target observations (airmass = 1.615) 
(see:
http://www.ls.eso.org/sci/facilities/lasilla/instruments/timmi/Reports/oschuetz/Projects/T2\_Extinc/TIMMI2\_extinc.html)
and Schuetz \& Sterzik (2004)).
Because of the different spectral types of the standard star (K2-III) and the target (solar-type), the color correction factor is
estimated to be 1--3~\% in the $N1$ filter (M{\"u}ller
et al., 2004).
The final result for the monochromatic flux density of C/2002~CE$_{10}$ at 8.7~$\mu$m is
0.50$\pm$0.05~Jy. The error accounts for the uncertainties in airmass
correction, as well as for the 1.6~\% stellar model uncertainty
(Cohen et al., 1999) and the spectral color correction.

\begin{figure*}%[!hbtp]
   \centering
%   \vspace{2mm}
%\includegraphics[width=6cm,angle=-90]%[width=6cm,angle=-90,clip]
\includegraphics[width=9cm]
{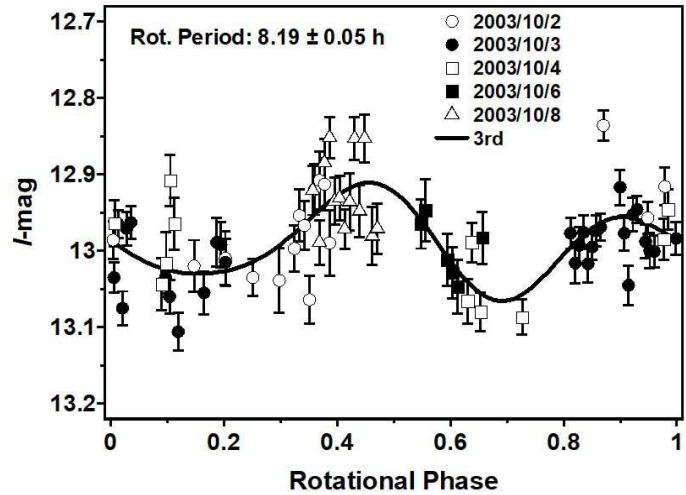}
%  \vspace{-5mm}
\caption{
Fitted lightcurve of C2002~CE$_{10}$ (solid line).
Differential photometric observations were performed from 
2003-Oct-2 to 2003-Oct-8
with the 1.05~m Schmidt telescope at the Kiso observatory in Japan. }
         \label{lightcurve}
\end{figure*}

   \begin{figure*}
   \centering
\includegraphics[angle=0,width=9cm]{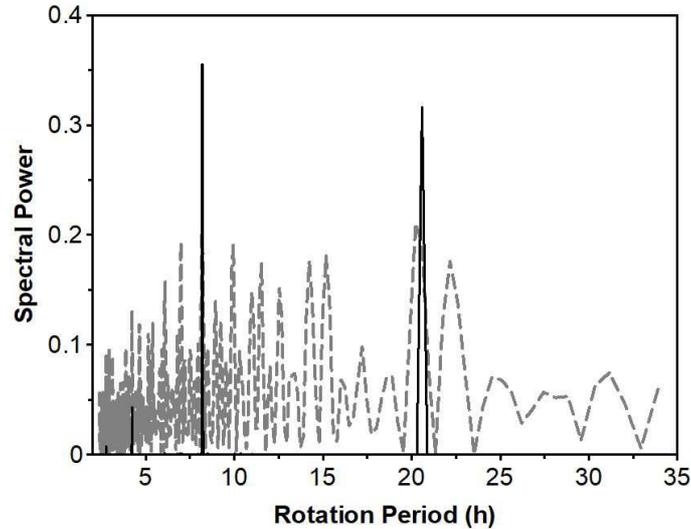}
\caption{
Analysis of the lightcurve--rotation period of C/2002~CE$_{10}$.
 The solid line shows the result for Window-CLEAN method, and the dashed-line
is for Lomb-Scargle spectral analysis. 
The uncertainty of the rotation period was calculated
considering the frequency interval (Fourier transform)
of Window-CLEAN around the highest peak of the obtained spectral power,
i.e. 0.04656~h. %(=0.00194~days).
%0.00194~day (= 0.05~h). 
The second highest peak is located at 20.58~h.
Note that the two highest peaks of Window-CLEAN coincide with those of
Lomb-Scargle, and the Lomb-Scargle scheme shows both peaks with 98~\%
significance level.
}
         \label{periodgram}
   \end{figure*}

\section{Results}
\label{results}

\subsection{Rotation period and body axis ratio}

Two independent methods were used for the periodogram analysis of
the $I$-filter lightcurve of C/2002~CE$_{10}$, e.g. Lomb's spectral analysis (Lomb 1976)
which is widely used as a standard method to calculate
the power spectra of unevenly spaced time-series data, and the Window-CLEAN method 
(Roberts et al. 1987) which was originally developed in radio astronomy 
for UV-plane image synthesis. 
More specifically, we applied the implementation of the time series 
analysis by Yoshida et al. (2016) used for asteroid lightcurve data.
This analysis approach was used previously to
determine the rotational period of asteroid 25143 Itokawa, which was later
confirmed by the Hayabusa space mission (Dermawan et al., 2002).
The two highest peaks of Window-CLEAN periodograms of the C/2002~CE$_{10}$ lightcurve data
coincide with those of Lomb-Scargle approach applied in parallel to the same 
dataset. 
The Lomb-Scargle scheme shows both peaks at
significance level of 98~\% (Fig.~\ref {periodgram}).
A rotation period of 8.19$\pm$0.05~h
%8.19~h (= 0.341~days) 
is considered the most probable
solution for C/2002~CE$_{10}$.
However, other solutions for the rotation period of the object
cannot completely ruled out based upon the dispersion inherent in the photometric data of 
C/2002~CE$_{10}$ (Fig.~\ref{lightcurve}).

Assuming that the lightcurve of C/2002~CE$_{10}$ are caused 
by the non-spherical shape of its nucleus alone (e.g. Sekiguchi et al., 2002),
a lower limit for the shape elongation, i.e. the $a / b$ axis ratio,
can be estimated as
$ a / b \geq %\gid 
10^{0.4 \, {\Delta} m}  = 1.2\pm0.1$, 
where ${\Delta} m$ is the peak-to-valley amplitude of the fitted lightcurve (in magnitude).

\subsection{Absolute Magnitude and Spectral Colors}

For the determination of the absolute magnitude of C/2002~CE$_{10}$
the analysis approach based upon 
the $HG$ system (Bowell et al., 1989) of the International Astronomical Union IAU is chosen:
\begin{equation}
V (1, 1, \alpha) = V (1,1,0) - \log
 \Bigl [ (1- G )  
\Phi_1 (\alpha) + G \Phi_2 (\alpha) \Bigr ]^{2 \over 5},
\label{Hapke}
\end{equation}
where $V (1, 1, \alpha)$ is the $V$-band magnitude reduced to unity Sun and Earth distances
$R_{\mathrm h}=\mit\Delta=$ 1~au at phase angle $\alpha$, and $V (1,1,0)$ is the
absolute magnitude $H$.
$G$ represents the slope parameter in $V$-band. 
$\Phi_1$ and $\Phi_2$ are describing the phase function and are
considered filter-independent.
Since for C/2002~CE$_{10}$ the phase angle coverage of our observations is not sufficient, we
applied a predetermined values of 0.15 as a default slope
parameter $G$ (e.g. Bowell et al., 1989).
The estimated absolute magnitude of C/2002~CE$_{10}$ in $V$-band is 
$H=13.08\pm0.07$~mag. This result is used together with the thermal flux of the
object in order to constrain the size and albedo of C/2002~CE$_{10}$ - see next section.  

\begin{table*}
\caption{Colors of  C/2002~CE$_{10}$ in magnitude}
\begin{tabular}{ccccc}
\hline
\hline
$B - V$ & $V - R$  & $V - I$ & $ V (1, 1, 0)^{\mathrm a}$ \\
\hline
0.734 $\pm$0.034 & 0.568 $\pm$0.039  & 1.075 $\pm$0.033 &
13.07$\pm$0.07 \\
\hline
\end{tabular}
%Times are given in hours and minutes,
\begin{list}{}{}
\item[$^{\mathrm{a}}$] $V (1, 1, 0) =H$ is the absolute magnitude in $V$-band
%\item[$^{\mathrm{b}}$] $G$ is the slope parameter in the IAU $HG$ system.
\end{list}
\label{optcolor} 
\end{table*}

   \begin{figure*}
   \centering
\includegraphics[angle=0,width=8cm]{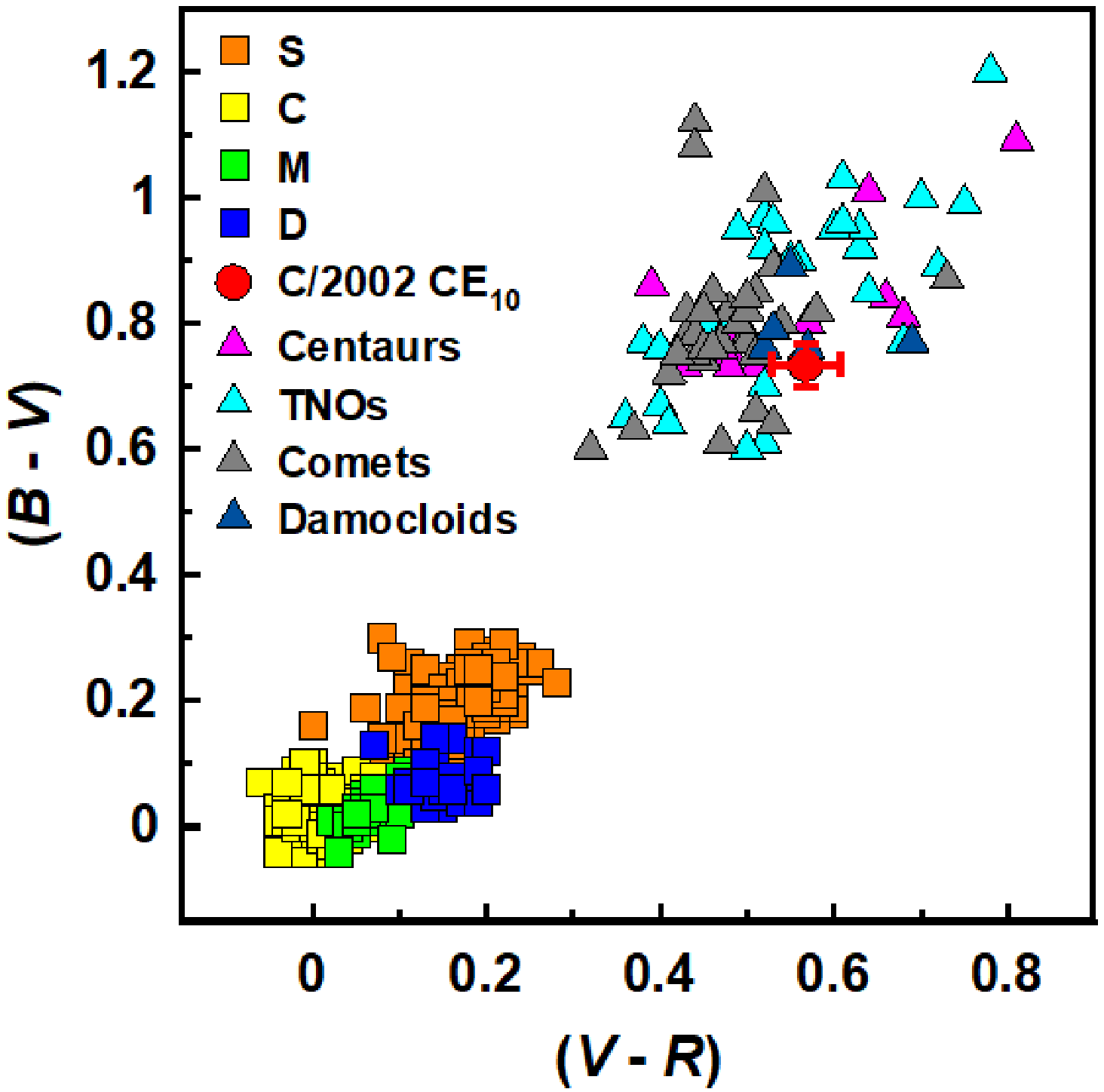}
\includegraphics[angle=0,width=8cm]{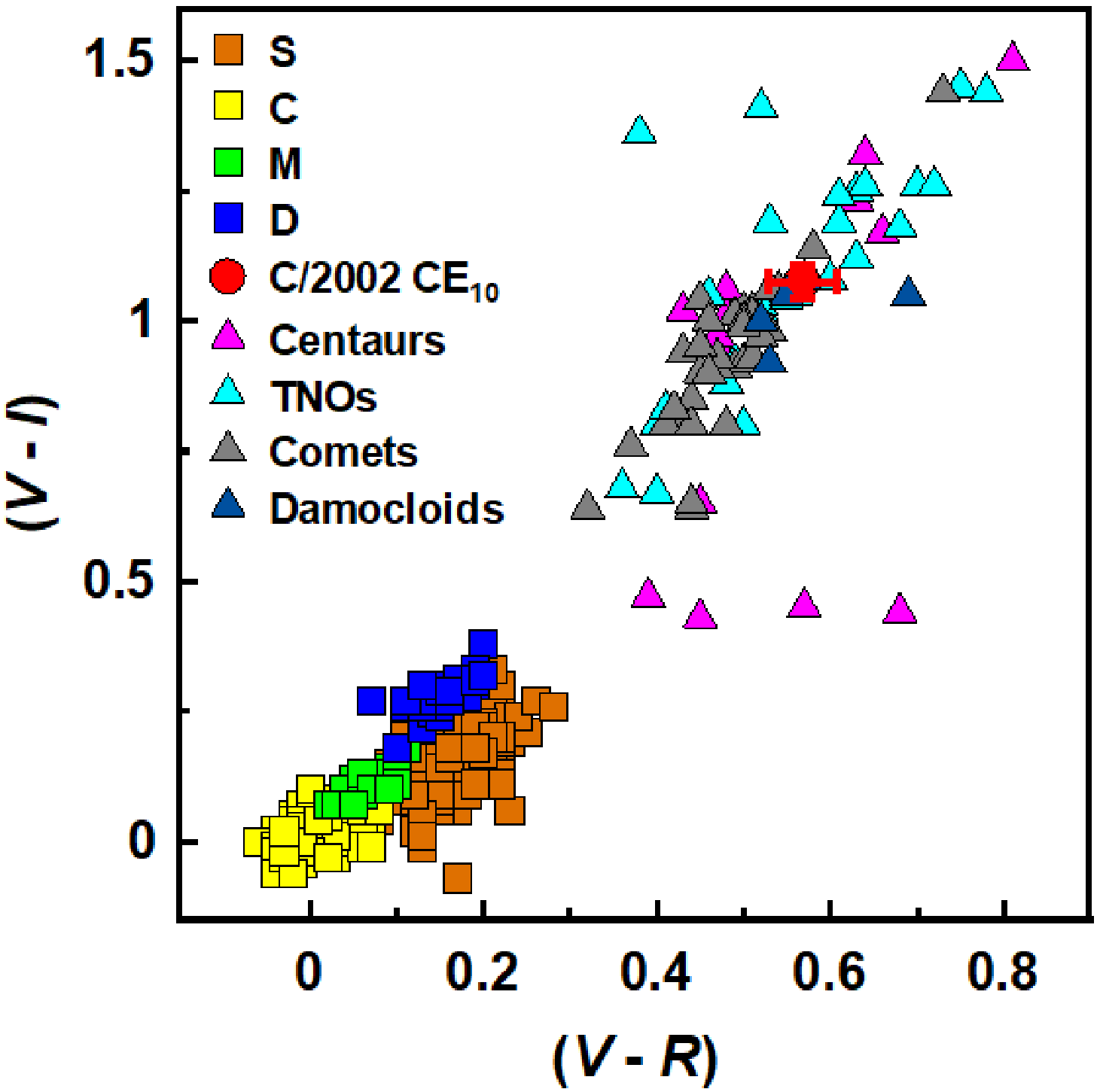}
\caption{
Color-color diagrams of different minor body groups including C/2002~CE$_{10}$.
The comet and Damocloids data were taken from 
Abell et al. (2003), Campins et al. (2006), Doressoundiram et al. (2007), Hicks and Bauer (2007),
Jewitt et al. (2009), Lamy and Toth (2009), Jewitt (2015).
The square symbols in Fig.~\ref{colorplot} denote the taxonomy
ranges of S-type, C-type, M-type and D-type asteroids, respectively,
from the eight-colors asteroid survey data by Zellner et al. (1985)
and from NASA's Planetary Data System: Small Bodies Node (https://pds.nasa.gov/).
The colors for TNOs are taken from Boehnhardt et al. (2003), Sheppard and Jewitt (2002),
Mueller et al. (2004), Peixinho et al. (2004), Doressoundiram et al. (2007),
Sheppard (2010), Perna et al. (2013) while the Centaurs data are taken from the papers of
Gutierrez et al. (2001), Bauer et al. (2003), Doressoundiram et al. (2007), 
Perna et al. (2013), Jewitt (2015). 
}
         \label{colorplot}
   \end{figure*}

The photometric results of C/2002~CE$_{10}$ are illustrated in terms of colors $B-V$, $V-R$, $V-I$ (see Tab.~\ref{optcolor})  
that are representing coarse measures of the global surface taxonomy of the object.
In Fig.~\ref{colorplot} the three broadband colors of C/2002~CE$_{10}$ (marked by circle with error bars) are plotted
together with the colors of asteroids (colored squares), comets, Transneptunian Objects (TNOs) and
Centaurs (colored triangles). 
%Fig.~\ref{colorplot} shows the surface color of C/2002~CE$_{10}$. 
The surface taxonomy of C/2002~CE$_{10}$ is clearly  
redder than and much beyond the colors of the different asteroid types. 
It falls well in the middle of the range for TNOs, Centaurs, and cometary nuclei.

\subsection {Thermal mid-IR}

 The thermal IR count rates of C/2002~CE$_{10}$ and of the standard star 
were measured from the respective $N$-band images using the
aperture photometry method. 
Using the determined thermal flux of the
standard star, the $N$-band flux of C/2002~CE$_{10}$ was determined to be 0.50 $\pm$0.05~Jy.
The error accounts for measurement uncertainties, 
the uncertainties in airmass correction, as well as for
the 1.6~\% stellar model uncertainty (Cohen et al. 1999) and
in the color correction. 

The thermal flux density of mid-IR radiation is given by
\begin{equation}
S_{\nu} = \pi \, \varepsilon \; \Bigl({{r_{\mathrm N}} \over {\mit
\Delta}} \Bigr)^2 \;
B_{\nu}( T),
\label{S} 
\end{equation}
where $r_{\mathrm N}$ is the radius of C/2002~CE$_{10}$ (in m),
$\varepsilon$ = 0.9 (Lebofsky et al., 1986) is the infrared emissivity, 
$\mit \Delta$ is the geocentric distance (in m) and
$B_{\nu}(T)$  is the Planck function for the effective surface temperature $T$.
The effective surface temperature of minor bodies
is determined from the energy balance at the surface (e.g. Lebofsky \& Spencer, 1989). 
However, it also implies knowledge or an estimation of the surface temperature
of the object. Since by measurements of a single thermal filter band, the surface temperature
distribution of a minor body cannot be estimated, a thermal model approximation
is used instead. Usually, this approach has to adopt additional values for
in principle unconstrained parameters like the chemical composition, material distribution (albedo map), 
emissivity, density (and porosity), heat conductivity and heat capacity of
the surface materials, of the rotation period and the orientation of the rotation axis with respect to the Sun. 

For our analysis of the TIMMI2 data of C/2002~CE$_{10}$, we applied the so called 
NEATM model (Near Earth Asteroid Thermal Model; Harrris, 1998) that is adopted
to a specific object group. The NEATM is modified from the Standard Thermal Model (STM) 
of Lebofsky et al. (1986) which is usually used for minor body applications
(e.g. Sekiguchi et al., 2003).

Fern{\' a}ndez et al. (2013) and Licandro et al. (2016) show that 
comets as well as asteroids in cometary orbits (including Damocloids)
present beaming parameters of $\eta=1.0\pm0.2$.
The bond albedo $A$ and 
geometric albedo are related by
$A = p \; q  = p_{\mathrm V} \; q$,
where $p$ and $p_{\mathrm V}$ are the bolometric geometric albedo and the
geometric albedo in $V$-band, respectively. $q$ is the bolometric
phase integral which in the $H$-$G$ system 
is derived from the slope parameter $G$, 
via $q = 0.290 + 0.684 \; G$ (Bowell et al., 1989).

\begin{figure*}
\centering
\includegraphics[angle=90,width=12cm,clip]{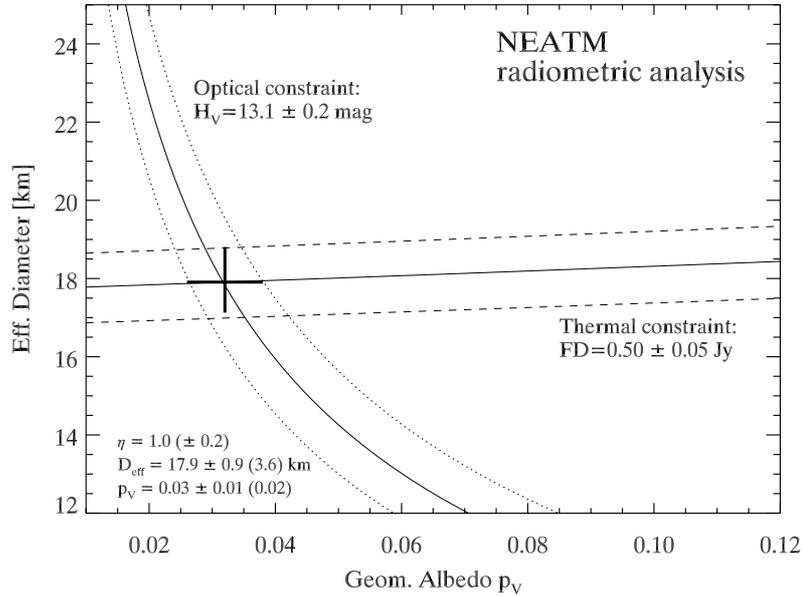}
\caption{
Size and albedo relationship of C/2002~CE$_{10}$.
The figure shows a graphical representation of the
    radiometric method: The solid (with error) curve
    represents the "optical constraint", based on
    the $H$-magnitude. The almost horizontal curve (with error) 
    represents the "thermal constraint" coming from the
    TIMMI2 measurements. The thermal flux 
    is tightly connected to the size of the object.%
}
\label{albedorad}
\end{figure*}

The absolute magnitude $H$, geometric albedo $p_{\mathrm V}$ and diameter $D$
(km) for asteroids are related by (e.g., Fowler \& Chillemi, 1986)
\begin{equation}
%%%%	D = 1.329 \times 10^{0.2 H} \; {p_{\mathrm V}}^{-0.5}
\log_{10}D = 3.1236 - 0.2~H - 0.5~\log_{10}(p_{\mathrm V}), 
\label{D}
\end{equation}

Fig.~\ref{albedorad} shows the diameter versus the geometric
albedo relationship for C/2002~CE$_{10}$ using the measurement results in the
visible and thermal wavelength range (including uncertainties).
In the figure  the optical constraint, representing the $H$ magnitude of the object 
in Eq.~(\ref{D}), is shown as decreasing curves.
The thermal constraint representing the IR flux of the object is
an almost horizontal curve (plus parallel lines indicating the uncertainty ranges).
The simultaneous solution of thermal and optical constraint in Eq.~\ref{S} and Eq.~\ref{D}, respectively,
is the the intersection of the respective curves. Thus, the best estimate for the
radius $r$ and albedo $p_{\rm V}$ of C/2002~CE$_{10}$ is found to be: $r=17.9\pm0.9$~km,
$p_{\rm V}=0.03\pm0.01$.

\section{Discussion}
\label{discussion}
The dynamical and physical characteristics of C/2002~CE$_{10}$ suggest several similarities
with minor bodies from the outer solar system: very low Tisserand parameter, red color, low albedo. 
The presence of a tail in 2002 indicates a link to the comets population, although it is not known whether it is
driven by repetitive and extended periods of cometary activity due to ice sublimation, or by a 
rare and singular activity event, caused for instance by an impact.
 
C/2002~CE$_{10}$ is in an orbit that crosses the orbital distances of the giant planet. For this
reason it is likely in a transitional orbit state, since it is exposed to encounters with the gas giants in the
planetary system and thus to gravitational scattering in the future and it was so in the past. Various reservoirs
of minor bodies can be considered as the original home
region of C/2002~CE$_{10}$: 
The Kuiper Belt and the Oort Cloud as main reservoirs for the short-periodic and long-periodic comets, respectively.
While the known physical characteristics (size, color, albedo, axis ratio) are no very conclusive (whether the object
has to be seen as short-periodic comet coming from the Kuiper Belt or as long-periodic comet from the Oort Cloud), 
the Tisserand parameter and the high inclination orbit of C/2002~CE$_{10}$ clearly favor an origin as for the 
long-periodic comets.

The Oort cloud objects (a$\sim$10,000~au) are thought to 
have originally formed in the giant planet region. As soon as the giant planets had formed, they were removed from 
that region in the formation disk of the solar system by gravitational interaction with these planets and got
stranded and dynamically thermalized in the Oort Cloud (e.g. Weissman et al. 2002). Inward scattering
by masses in the closer and further distance of the solar system in the galaxy brings Oort Cloud comets back
into the range of the planets. 

The non-detection or the very weak level of cometary activity during the perihelion arc of the orbit
that C/2002~CE$_{10}$ passed through in 2002 to 2003 with solar distances well within the water sublimation limit, 
may indicate a rather exhausted state of the nucleus activity. So, C/2002~CE$_{10}$ can be seen as a member of the group of Damocloids. 
Its surface color is in good agreement with the average color of Damocloids:
$B-V=0.80\pm0.02$, $V-R=0.51\pm0.02$, $R-I=0.47\pm0.02$, $B-R=1.31\pm0.02$ 
(Jewitt, 2015). Jewitt (2005) reported the lack of ultrared matter in Damocloids as for instance seen in the
population of the dynamically cold Classical Disk Objects in the Kuiper Belt (Hainaut et al., 2012). 
The estimated size of C/2002~CE$_{10}$ 
of $17.9\pm0.9$~km is in the middle (a little bigger) of the size range obtained for Damocloids
using WISE data (Licandro et al., 2016). The derived radiometric albedo of $0.03\pm0.01$ 
(or 0.02 if the an $\eta$ range 0.8--1.2 is considered) agrees with the albedo range of 0.02--0.06 
for cometary nuclei (see for instance Campins and Fern{\' a}ndez 2002, Lamy et al. 2004).
It is close to the mean visible geometric albedo of 0.04 for Jupiter-family
comets obtained by Fern{\' a}ndez et al. (2013), and also close to the mean geometric albedo 
of $0.05\pm0.02$ found by Licandro et al. (2016) for asteroids in cometary orbits. On the other side, the albedo of C/2002~CE$_{10}$
represents the lower limit of the albedos found for Kuiper Belt objects (Lacerda et al. 2014) and is clearly below
that of several dynamical sub-populations therein. However, its surface colors would comply with these, while its 
retrograde orbit may not be easily accomplished through gravitational scattering from that region by a giant planet.

Comparing C/2002~CE$_{10}$ with short/long-periodic comets requires an explanation of its low level or absence of activity 
during perihelion: Exhaustion of the sublimating ices or coverage of the surface by thick layers of regolith from 
previous activity cycles that may prevent the solar heat wave to reach the still present ice reservoirs underneath.
The fraction of active regions on cometary surfaces with volatile ices is generally small and most of the
activity originates from sub-surface layers (e.g. Keller et al. 1986). Recently, the ROSETTA mission has obtained
a significant number of detailed images of 67P/Churyumov-Gerasimenko that reveal a lack of distinct active region 
with exposed fresh ice chunks (Thomas et al., 2015). The mission has also demonstrated the existence of meter-thick
regolith on the nucleus surface that represents ballistic fall-back material from cometary activity.

The "Grand Tack" model showed that the giant planets scattered inner solar system material outward
during their inward migration, and vice versa, they scattered icy planetesimals into the inner solar system
during their outward migration (Walsh et al, 2011). 
The surface color of C/2002~CE$_{10}$ is much redder 
than that of asteroids %and the geometric albedo is very dark 
and  perhaps its extremely low Tisserand parameter and retrograde orbit
which make on origin of the object in the main asteroid belt unlikely.
However, a scenario of an inward scattered and stranded object from the outer solar system
may be a valid explanation for C/2002~CE$_{10}$. 

Recently inactive minor body with hyperbolic
eccentricity, 1I/2017 U1($^\prime$Oumuamua)
was discovered
and its physical properties are studied (Meech et al, 2017).
It can be either from interstellar originally 
or from other planetary systems, or from our Oort cloud 
as a result of multiple scattering due to stellar encounters.
The investigation of the evolution of the Oort cloud 
may give a hint to understand 
the relationship among such outer rocky minor bodies.

\section*{Acknowledgments}

We are grateful for the recommendations and suggestions
to this manuscript made by Olivier Hainaut, Henry Hsieh and
one anonymous reviewer.
This work is based on
observations collected at the European Organization for
Astronomical Research in the Southern Hemisphere ESO under
programme 60.A-9126(F).

%%


\bibliographystyle{elsarticle-num}
\bibliography{<your-bib-database>}






% \end{thebibliography}

\end{document}